\documentclass[aps,prl,showpacs,twocolumn]{revtex4}
\usepackage{amsmath,graphicx}

\newcommand{\ra}{\rangle}
\newcommand{\la}{\langle}
\newcommand{\ain}{\hat{a}_{{\rm in}}}
\newcommand{\aout}{\hat{a}_{{\rm out}}}
\newcommand{\bin}{\hat{b}_{{\rm in}}}
\newcommand{\bout}{\hat{b}_{{\rm out}}}
\newcommand{\hc}{\hat{c}}
\newcommand{\kb}{\kappa_b}
\newcommand{\ka}{\kappa_a}
\newcommand{\hT}{\hat{\theta}}
\newcommand{\hN}{\hat{N}}

\begin{document}
\title{Measurement induced entanglement and quantum computation with atoms 
in optical cavities}

\author{Anders S. S\o rensen$^1$ and Klaus M\o lmer$^2$}
\affiliation{$^1$ITAMP, Harvard-Smithsonian Center for Astrophysics,
Cambridge Massachusetts 02138\\
$^2$Danish National Research Foundation Center for Quantum 
Optics \\ Department of Physics and Astronomy, University of Aarhus, 
DK-8000 Aarhus C, Denmark}  

\begin{abstract}
  We propose a method to prepare entangled states and  implement quantum
  computation with atoms in optical cavities. The internal state of the 
  atoms are entangled by
  a measurement of the  phase of light transmitted through the cavity.
  By repeated measurements an entangled state is created with certainty, 
  and this
  entanglement can be used to implement gates on qubits which are stored in
  different internal degrees of freedom of the atoms. This method, based
  on measurement induced dynamics, has a higher 
  fidelity than schemes making use of controlled unitary dynamics.
\end{abstract}


\maketitle

An essential ingredient in the construction of a quantum computer is
the ability to entangle the qubits in the
computer. In most proposals for quantum computation this entanglement
is created by a controlled interaction between the quantum
systems which store the quantum information \cite{fortschritte}. 
An alternative strategy to
create entanglement is to perform a measurement which
projects the system into an entangled state \cite{cabrillo,
brian}, and based on this 
principle a quantum computer
using linear optics, single photon sources, and single photon
detectors has recently been proposed \cite{linear}. Here we present a
similar proposal for measurement induced entanglement and quantum
computation
on atoms in  optical cavities, which only uses coherent light
sources and homodyne detection, and we show that high
fidelity operation can be achieved 
with much weaker requirements for the cavity and atomic parameters than 
in schemes which rely on a controlled interaction. 

Several schemes for measurement induced entanglement have been
proposed for atoms in optical cavities
\cite{bose,plenio,hong,duan,plenioto}, but compared to these 
schemes our proposal has the advantage that entanglement can be
produced with certainty even in situations with finite detector
efficiency. Furthermore, our procedures are similar to methods
already in use to monitor atoms in cavities \cite{hood}, and we
therefore believe that the current proposal should be simpler to
implement. Indeed the atom counting procedure in the experiment 
in Ref.\ \cite{mckeever}
could be sufficient to implement this scheme. 

\begin{figure}[tb]
  \centering
  \includegraphics[width=8cm]{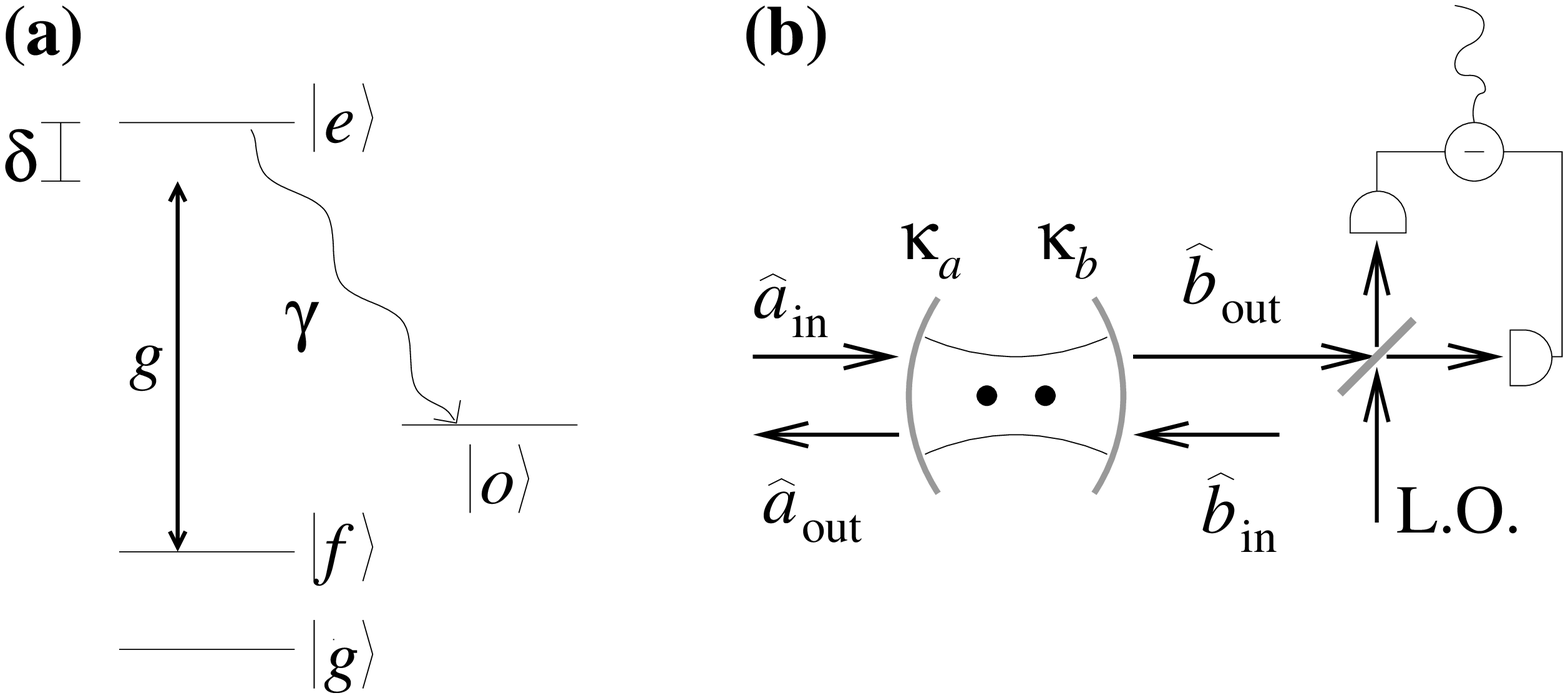}
  \caption{Energy levels and experimental setup of the atoms. (a) 
  The atoms 
  have two groundstates $|g\ra$ and $|f\ra$, and the cavity
  couples the state $|f\ra$ to the excited state $|e\ra$ with coupling
  strength $g$ and detuning $\delta$. The excited state $|e\ra$
  decays to an auxiliary state $|o\ra$ with a decay rate $\gamma$.
  (b) Two
  atoms are located inside a cavity and light is
  shined into the cavity in the mode described by $\ain$.
  After mixing with 
  a local oscillator (L.O.), the transmitted light is measured by
  homodyne detection.}
  \label{fig:setup}
\end{figure}

We consider atoms
with two stable ground states $|g\ra$ and $|f\ra$ and an excited state
$|e\ra$ as shown in Fig.\ \ref{fig:setup} (a). Both atoms are initially
prepared in an equal superposition of the two ground states,
$(|g\rangle+|f\rangle)/\sqrt{2}$.
% so that
% their combined state is
%\begin{equation}
%  \label{eq:inistate}
%  \frac{1}{2}|gg\ra +\frac{1}{\sqrt{2}}\frac{|gf\ra
%  +|fg\ra}{\sqrt{2}}+\frac{1}{2} |ff\ra.
%\end{equation}
By performing a Quantum Non-Demolition (QND) measurement 
which measures  $N$, the number of atoms in state $|f\ra$, the state vector
is projected into the maximally
entangled state $|\Psi_{{\rm EPR}}\ra=(|gf\ra+|fg\ra)/\sqrt{2}$ provided we
get the outcome $N=1$. If the outcomes $N=0$ or
$N=2$ are achieved, the resulting state is $|gg\ra$ or
$|ff\ra$, so that the initial state can be
prepared once again, and we can repeat the
QND-measurement until the desired entangled state
is produced. On average two measurements suffice to produce
the state.
%Since
%the probability for  
%the preparation of the entangled state is 1/2 we can prepare an
%entangled state  
%by performing the detection only two times (on average). 
With $M$ atoms in the cavity,
the scheme can be extended to the generation of multiparticle
entangled states where the population of the $|f\rangle$
state is distributed symmetrically
on all atoms. If, e.g., $M=3$ we may produce
with certainty the W-states \cite{w-state}
$(|ffg\ra+|fgf\ra+|gff\ra)/\sqrt{3}$    
by performing  the detection  $4/3$ times on average. Also, one could
entangle a small subset of the atoms by leaving all other
atoms in $|g\ra$ so that they do not contribute to $N$. 

The implementation of the QND detection is a modification of
a scheme presented in Ref.\ \cite{reflection} where QND detection was
achieved by measuring single photons reflected from a cavity. To generate
entangled states with a high fidelity this method is
undesirable because it is very sensitive to imperfections in the
cavity and to imperfect mode matching.
Instead we propose to measure the light 
transmitted through the cavity with homodyne detection.
The proposed experimental setup is
shown in Fig.\ \ref{fig:setup} (b). The atoms are trapped inside the
cavity, and we consider a single field mode described by the
annihilation operator $\hc$. Photons in the cavity can decay through two
leaky mirrors with decay rates $\kappa_a$ and $\kappa_b$, and the
incoming and outgoing fields at the mirror with decay rate $\kappa_a$
($\kappa_b$) are described by $\ain$ and $\aout$ ($\bin$ and $\bout$).
Light is shined into the cavity in the
$\ain$ mode, and the transmitted light in the $\bout$ mode
is measured by balanced homodyne detection. In the cavity the
light couples the state $|f\ra$ of the $k$th atom to the excited
state $|e\ra$ with a coupling constant $g_k$. We mainly
consider light far from resonance with
the atoms, where the interaction 
changes the phase of the field by an angle proportional to $N$.
 By measuring the phase of the transmitted
light with homodyne detection, we thus obtain the desired QND
detection of $N$.  Spontaneous emission from the excited state limits
the amount of light we can send through the cavity, and below we
evaluate the fidelity $F$, obtainable for a given set of cavity
parameters ($F=\la \Psi_{{\rm EPR}}|\rho |\Psi_{{\rm EPR}}\ra$, where
$\rho$ is the atomic density matrix). The excited state
$|e\ra$ decays with a rate $\gamma$, and for simplicity we assume that the
atoms end up in some other state $|o\ra $ after the decay. Because
these states have a
vanishing overlap with the desired state our results will lead to a lower
fidelity than if we had assumed that the atoms decay back to the state
$|f\ra$, but it does not affect the scaling with cavity parameters,
which is our main interest  
here. 

The interaction of the cavity light with the atoms is described by the
Hamiltonian
\begin{equation}
  \label{eq:ham}
  H=\sum_k g_k |e\ra\la f|_k \hc+g_k^* \hc^\dagger |f\ra\la e|_k +\delta
  |e\ra\la e|_k,
\end{equation}
where $\delta$ is the detuning of the atoms from the cavity
resonance,
and where the sum is over all atoms in the cavity. 
The decay of the $k$th atom is described by a Lindblad relaxation
operator $d_k=\sqrt{\gamma} |o\ra\la e|_k$, and the input/output
relations are given by $ d \hc/ dt=-i [ \hc, H]-(\kappa/2) \hc+
\sqrt{\kappa_a}\ain +\sqrt{\kappa_b}\bin+\sqrt{\kappa'}\hat{F}$,
$\aout=\ain-\sqrt{\kappa_a}\hc$, and $\bout=\bin-\sqrt{\kappa_b}\hc$,
where $\kappa=\ka+\kb+\kappa'$ is the total decay rate, and
$\kappa'$ and $\hat{F}$ is the decay rate and noise operator 
associated with cavity loss to other modes than the two modes
considered. For convenience $\hc$ has the standard normalization for a
single mode $[\hc,\hc^\dagger]=1$, whereas the normalization for the
free fields are such that, e.g.,
$[\ain(t),\ain(t')^\dagger]=\delta(t-t')$. 
We assume that the atoms are separated by more than the 
resonant optical wavelength, and we hence ignore
the dipole-dipole interaction between the atoms.
We solve the equations of motion by assuming that the fields are
sufficiently weak that 
only the lowest order terms in $\hc$ are important. Note that this
limits the total number of photon in the cavity at any time, but 
the total number of photons in a pulse may still be large.  
By taking the Fourier transform we find the
cavity field to be given by \cite{reflection}
\begin{equation}
  \label{eq:comega}
  \hc(\omega)=\frac{\sqrt{\ka}\ain(\omega)}
    {\frac{\kappa}{2}-i\omega+
    \frac{g^2\hN}{\gamma/2+i(\delta-\omega)}}+{\rm noise},
\end{equation}
where we have introduced the
operator $\hN=\sum_k |f\ra\la f|_k(t=0)$ which counts the number of atoms in
state $|f\ra$ before the pulse, and we have assumed that the magnitude 
of all coupling
constants are identical $|g_k|^2=g^2$.
Using Eq.\ (\ref{eq:comega}) we find an
expression for the
transmitted light 
\begin{equation}
  \label{eq:outfield}
  \bout(\omega)=\frac{-\sqrt{\ka\kb}}{\frac{\kappa}{2}-i\omega+
  \frac{g^2\hN}{\gamma/2+i(\delta-\omega)}  }\ain(\omega) + {\rm noise}. 
\end{equation}

We determine the Heisenberg equations of motion for the atoms from
Eq.\ (\ref{eq:ham})  
and the relaxation operators and solve them by assuming that the
atomic ground state  
operators vary only little on the time scale $1/\gamma$. Here we shall
only need     
the projection operator onto the $|f\ra$ state, which
after the interaction with the light pulse $(t=\infty)$ is given by
$|f\ra\la f|(t=\infty) 
= : |f\ra \la f|(t=0) \exp(-\hat{\theta}):$, 
where $:\quad :$ denotes normal
ordering, and where we have introduced the scattering
probability operator 
\begin{equation}
  \label{eq:theta}
  \hT=\gamma\int d\omega \frac{g^2
  } {\frac{\gamma^2}{4}+(\delta-\omega)^2}\hc(\omega)^\dagger\hc(\omega).
\end{equation}

    The incoming light is assumed
  to be in a coherent state with frequency $\omega$ and
  since  Eq.\
  (\ref{eq:outfield}) is linear 
%for any number of atoms $N$ in state $|f\ra$,
 the outgoing field will also be in a
  coherent state  $|\beta_N \ra$. 
  Using Eqs.  (\ref{eq:comega})-(\ref{eq:theta})  we see that the
  difference between  
 the coherent state  output of an empty
  cavity and that of a cavity with $N$ atoms can be expressed as 
  \begin{equation}
    \label{eq:distone}
    |\beta_N-\beta_0|^2=N^2 \frac{g^2\kappa_b}{\gamma(\frac{\kappa^2}{4}
    +\omega^2)} \theta_N,
  \end{equation}
 The scattering probability per atom $\theta_N$ [obtained by inserting
 Eq.\ (\ref{eq:comega}) in Eq. (\ref{eq:theta})]  depends on the 
  number of atoms, because the field amplitude inside the cavity depends 
  on $N$.   From
  Eq.\ (\ref{eq:distone}) we see that the best signal
  to
  noise ratio  ($|\beta_N-\beta_0|^2/\theta_N$) is obtained by use of
  light which is resonant with the 
  cavity, $\omega=0$. For resonant light, the transmittance of the
  output mirror should match all
  other losses  $\kb=\ka+\kappa'=\kappa/2$, and the QND detection then only
  depends on the cavity parameters through the combination
   $g^2/\kappa\gamma$. In particular,
   $|\beta_N-\beta_0|^2=N^2\theta_N g^2/\kappa\gamma$. If $\kb$ deviates from
  $\kappa/2$ the effective $g^2/\kappa\gamma$ is changed by
   $2\kb/\kappa$. If transmitted photons are lost, or if they are only
detected with a probability $\eta$, the
  effective $g^2/\kappa\gamma$ is reduced by this factor.

If we need to distinguish an empty cavity from a cavity with a
certain number of atoms, the signal to noise ratio is  independent
  of the detuning of the cavity from the atomic resonance, but here we
   concentrate on the far detuned case  $\delta\gg
  g^2/\kappa$, $ \gamma$,  where the $N=1$ and $N=2$ components are 
  easier to distinguish.
  In this limit the intra cavity field and therefore
  the scattering probability $\theta_N$ is independent of $N$ and we
  replace $\theta_N$ by the same $\theta$ for all $N$.  If the light
  is on resonance with the cavity $\omega=0$, and far detuned from
  the atomic resonance, we see from Eq.\ (\ref{eq:outfield})
that the interaction with the atoms  changes the
phase of the field by an angle proportional to $N$. By adjusting the
phase of the local oscillator we can configure the
measurement so that the photocurrent
difference, integrated over the duration of the pulse, can be normalized to
a dimensionless field quadrature operator $x$ which, in turn, is
proportional to the phase change imposed by the atoms and, hence, it
provides a QND detection of $N$. If the atoms
are in the $N=0$ state, light is transmitted through the cavity
without being affected by the atoms, and the homodyne 
measurement provides
a value for $x$ in accordance with a Gaussian
probability distribution $p_0(x)=\exp(-x^2)/\sqrt{\pi}$ with a
variance $(\Delta x)^2=1/2$ due to vacuum noise in the coherent
state. If $N=1$ or 2, Eq.\ (\ref{eq:distone}) predicts  the
Gaussian to be displaced so that it is centered around
$x_N=2N\sqrt{\theta g^2/\kappa\gamma}$. 

So far we have ignored that the atoms
are slowly pumped out of the state $|f\ra$ due to decay of the upper state, 
and this changes the distribution. 
If we start with  $N$ atoms in state $|f\ra$,
the probability distribution $p_N(x)$
can be written as a sum of a part $p_{{\rm ND},N}(x)$ where there has been
no atomic decay and a part $p_{{\rm D},N}(x)$ where at 
least one of the atoms have decayed:
 $p_N(x)=p_{{\rm ND},N}(x)+p_{{\rm D},N}(x)$.
With no decay Eq.\ (\ref{eq:distone}) is valid and we have $p_{{\rm
ND},N}(x)=\exp(-N\theta)p_0(x-x_N)$. If there is a decay, Eq.\
(\ref{eq:outfield})  is valid until the time of the decay and we find
 $p_{{\rm D},1}=\int_0 ^\theta d\theta'
\exp(-\theta') p_0(x-x_1\theta'/\theta)$ and 
$p_{{\rm D},2}=2\int_0 ^\theta d\theta'
\exp(-\theta-\theta') p_0(x-x_1(\theta'+\theta)/\theta)+\int_0 ^\theta
d\theta'\int_0 ^\theta d\theta'' 
\exp(-\theta'-\theta'') p_0(x-x_1(\theta'+\theta'')/\theta) $.
If the homodyne measurement produces an outcome
 $\tilde{x}$ in a region $x_a<\tilde{x}<x_b$ around  $x_1$,  
 the atoms are most likely in the desired 
 entangled state and the measurement is successful. The probability for this
measurement outcome is 
$P_s=\int_{x_a}^{x_b} d\tilde{x} P_{\rm tot}(\tilde{x})$, 
where the total probability distribution for $\tilde{x}$  is given by
 $P_{{\rm tot}}(\tilde{x})= p_0(\tilde{x})/4+p_1(\tilde{x})/2
  +p_2(\tilde{x})/4$.

\begin{figure}[b]
  \centering
  \includegraphics[width=8cm]{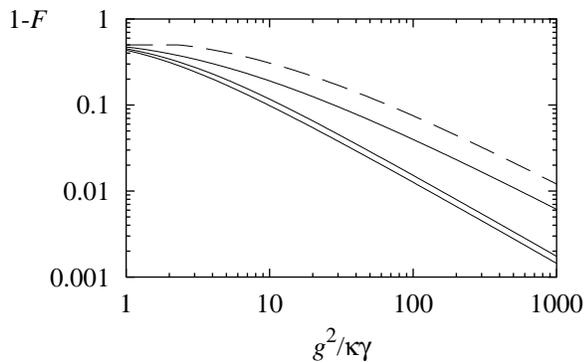}
  \caption{Fidelity of the produced entangled state. Starting from
  below, the full curves
  shows the obtainable error probability $1-F$ for $P_s=0.1\%$, 30\% and
  50\%.  The dashed curve is an upper bound on the total error probability
when the scheme is repeated until
  it is successful.}
  \label{fig:fidelity}
\end{figure}

The fidelity
of the entangled state vanishes if one of the atoms has
decayed to a different subspace, and the fidelity of the state produced is 
therefore given by the conditional probability of starting in the state
with $N=1$ and 
not having  decayed. By averaging over the interval of 
accepted values we get the fidelity, 
$F=\int_{x_a}^{x_b} d\tilde{x} p_{{\rm ND},1}(\tilde{x})/2/P_s$. 

For a given value of $g^2/\kappa\gamma$ the success probability and
  fidelity depend on $\theta$, $x_a$ and $x_b$. To get a
  high fidelity one must choose a small interval
  around $x_1$ because it minimizes the  contribution from  $p_0$
  and $p_2$, but a small interval also reduces the success probability. In
  Fig.\ \ref{fig:fidelity} we show the smallest error 
 probability $1-F$ achievable as a function
  of $g^2/\kappa\gamma$, when we vary $x_a$ and $\theta$ and
  adjust $x_b$ to retain a fixed value of $P_s$. In the figure the
  full curves show $1-F$
  for $P_s=0.1\%$, $P_s=30\%$ and $P_s=50\%$.

In Fig.\ \ref{fig:fidelity} we observe  that the fidelity
scales as 
\begin{equation}
  \label{eq:goodscaling}
  1-F\sim
  \frac{\kappa\gamma}{g^2} \log{\left(\frac{g^2}{\kappa\gamma}\right)}.
\end{equation}
This behavior can be understood from Eq.\ (\ref{eq:distone}): with a fixed
distance $|\beta_1-\beta_0|=\sqrt{2} x_1$, the error due
to scattering decreases as $\kappa\gamma/g^2$, but due to the overlap of the
Gaussians
there is a probability of accepting a wrong value for $N$,
which will dominate over the error due to scattering if we keep the
distance fixed and increase $g^2/\kappa\gamma$. Because the overlap of
the Gaussians decreases exponentially with the square of the distance, optimum
fidelity is obtained if the distance grows 
logarithmically with $g^2/\kappa\gamma$ and this gives the scaling in Eq.\
(\ref{eq:goodscaling}).

The scaling (\ref{eq:goodscaling}) is
valid up to a success rate $P_s=50\%$, where $P_{{\rm tot}}(\tilde{x})$ 
consists of three well separated
Gaussians, so that even when the scheme is not successful we know in
which state 
($N=0$ or $N=2$) the atoms are left.  From these states we can
deterministically produce the initial product state, 
and we can repeat the measurement procedure until it
is successful. 
A lower limit on the fidelity with this procedure can be shown to be
 $F>\int_{x_a}^{x_b} d\tilde{x} p_{{\rm
ND},1}(\tilde{x}) /2/(1-\int_{-\infty}^{x_a}d\tilde{x}
p_0(\tilde{x})/4-\int_{x_b}^\infty d\tilde{x} p_{{\rm 
ND},2}(\tilde{x})/4)$. This lower limit is shown with a dashed
line in Fig.\  
\ref{fig:fidelity}, and the scaling still follows
(\ref{eq:goodscaling}).  

The deterministic
entanglement scheme can be used to
implement gates between qubits. Suppose that the levels $|f\ra$ and
$|g\ra$ are both degenerate with two substates.
For convenience the four states 
$|g_0\rangle,\ |g_1\rangle,\ |f_0\rangle,\ |f_1\rangle$
can be described as tensor
products of the 'level' degree of freedom ($|g\ra$ and $|f\ra$) and
'qubit' degree of freedom ($|0\ra$ and $|1\ra$). (This would be
the natural representation if $|0\ra$ and $|1\ra$ represent, e.g.,
different motional state, but they may also be
different internal states in the atoms, e.g., states with different magnetic
quantum numbers). The qubits are initially stored in the states $|g_0\ra$ and
$|g_1\ra$ and, we assume that
the interactions during the entanglement preparation are symmetric with 
respect to the qubit
degrees of freedom, so that the qubits are decoupled from the
preparation of the  entangled state of the levels, i.e, we prepare the state
$|\Phi_q\ra\otimes(|gf\ra+|fg\ra )/\sqrt{2}$, where
$|\Phi_q\ra$ is the initial state of the qubits. 
%(Note, that if there
%are more than two atoms in the cavity one can 
%select the two atoms participating in the gate be leaving all other
%atoms in $|g\ra$ so that they never interact with the cavity). 
The entanglement in the levels $|f\ra$ and $|g\ra$ can now be used to
implement a {\sc control-not} operation on the qubits as 
discussed in Ref.\ \cite{cnot}: we interchange states 
$|f_1\ra$ and $|g_1\ra$ in the control atom, 
and then measure if this atom is in the $|f\ra$ or the  
$|g\ra$ level. If the measurement
outcome is $|g\ra$ ($|f\ra$) the target atom is known to be in $|g\ra$
($|f\ra$) if the 
control atom is in $|1\ra$, and the {\sc control-not} operation amounts to
interchanging the states $|g_0\ra$ and $|g_1\ra$ ($|f_0\ra$ and
$|f_1\ra$) in the target atom. 
To complete the gate  a $\pi/2$ pulse is applied
between the levels $|f\ra$ and $|g\ra$, and a QND-measurement of the
level of the target atom, followed by single atom transitions,
conditioned on the 
outcome of the measurement leaves the system in 
$|\tilde{\Phi}_q\ra\otimes|gg\ra$, where $|\tilde{\Phi}_q\ra$ is the desired
outcome of the {\sc control-not} operation on $|\Phi_q\ra$.
The single atom measurements required for the gate operation can 
be done by homodyne detection of light
transmitted through the cavity, and the procedure is
accomplished with a fidelity scaling as in Eq.\ (\ref{eq:goodscaling}).

A different strategy for quantum computation in optical cavities is to
construct a 
controlled interaction between the atoms and the cavity
field \cite{pellizzari,domokos,zheng}: If an atom emits a photon into
the cavity mode this photon 
can be absorbed by another atom, and this constitutes an interaction
which can be used to engineer quantum gates. 
With controlled interactions, a first estimate of the error
probability $\epsilon$ in 
a single gate  is given by
$\epsilon\sim \mathrm{max}\{\gamma,\kappa\}/g$, but this can be
improved  by use of  Raman transitions or a large
detuning between the atoms and  the cavity mode. If the optical
transition is replaced
by a Raman transition where a classical beam with Rabi frequency
$\Omega$ and detuning $\Delta$ induces a transition between two
ground states and the simultaneous creation/annihilation  of a photon in 
the cavity, the effective
coupling constant is $g_{{\rm eff}} \sim g \Omega/\Delta$ and the
effective decay rate is $\gamma_{{\rm eff}}\sim \gamma
\Omega^2/\Delta^2$. By decreasing $\Omega/\Delta$ we can thus reduce
the error probability $\gamma_{{\rm eff}}/g_{{\rm eff}}$ due to
spontaneous emission. Similarly it has also been
proposed to reduce the effect of cavity decay by using transitions which
are detuned from the cavity \cite{zheng}. If the Raman transitions are
detuned by 
an amount $\delta$ from the cavity mode the effective coupling
constant between the atoms is $\chi\sim  g_{{\rm eff}}^2/\delta$
and the leakage rate is $\kappa_{{\rm eff}}\sim \kappa  g_{{\rm
eff}}^2/\delta^2$, so that the leakage can be reduced by
making $\delta\gg\kappa$. But, detuning from the cavity mode increases
the effect of spontaneous emission because the gate takes longer
time. Since the gate duration is  $t\sim 1/\chi$ the 
probability for
spontaneous emission is $\epsilon_\gamma\sim
\gamma\delta/g^2$ and the probability for a cavity decay is $\epsilon_\kappa
\sim \kappa/\delta$. The minimum error probability
$1-F=\epsilon_{\gamma} 
+\epsilon_{\kappa}$ is then found to scale as 
\begin{equation}
  \label{eq:badscaling}
  1-F\sim \sqrt{\frac{\kappa\gamma}{g^2}}.
\end{equation}
 Although Eq.\ (\ref{eq:badscaling}) has
 been derived 
 for a particular setup, we believe the scaling in Eq.\
 (\ref{eq:badscaling}) is characteristic for all existing 
 proposals which use a
 controlled interaction between the atoms. (Also the scheme in the
 Ref. \cite{beige} has the scaling (\ref{eq:badscaling}) if the
 photodetectors have finite efficiency, or if the scheme is required to
 implement gates deterministically).
 %A better scaling $\epsilon\sim \gamma\kappa/g^2$
 %has been predicted \cite{law} for the production of a Fock state in
 %a cavity,  
 %but this results assumes that one of the mirrors 
 %has a vanishing decay rate $\kappa'=0$. If the mirror has a finite
 %decay rate, 
 %the error due to photons leaving in the wrong direction is
 %$\kappa'/\kappa$ and by optimizing the reflectivity of the mirror with
 %the largest transmittance we recover the result in Eq.\
 %(\ref{eq:badscaling}) where $\kappa$ is the decay rate of the mirror with
 %high reflectivity. 
The fundamental problem is that the atomic decay and the cavity loss
provide two sources of errors which combine to give the scaling in
Eq. (\ref{eq:badscaling}).
For a given 
cavity, i.e., for given parameters $g$,
$\kappa$, and $\gamma$, 
we obtain the better scaling presented in Eq.\ (\ref{eq:goodscaling})
by using a 
scheme where classical light is injected into the cavity and where
leakage of a single photon is not a critical event. 

In conclusion, we have presented a scheme to entangle atoms and
implement gates between qubits which are stored in the atoms. The
error probability scales more favorably
than it does in schemes which use
controlled interactions and unitary dynamics, and the present scheme
is therefore advantageous if a high fidelity is required. The best optical
cavities currently reach $g^2/\kappa\gamma\sim 100$ \cite{hood} so that Eq.\
(\ref{eq:badscaling}) predicts $F\sim 90\%$ which is less than
the value in Fig.\ \ref{fig:fidelity}. More importantly, to decrease
the error rate in the measurement induced scheme, e.g.,  
by a given factor we only need to improve the cavity 
finesse, which is proportional to $g^2/\kappa\gamma$, by the same
factor. With the scaling in Eq.\ (\ref{eq:badscaling})  the cavity
finesse should be increased by the same factor squared. 

This work was supported by the Danish Natural Science Research Council
and by the Natural Science Foundation through its grant to ITAMP.

\end{document}